\documentclass[]{JHEP3}

\usepackage{amsmath}
\usepackage{esint}
\usepackage{epsfig}
\usepackage{slashed}
\usepackage{rotating}
\usepackage{nicefrac}
\usepackage{textcomp}
\usepackage{subfigure}

\newcommand{\na}{\widetilde{\chi}_1^0}
\newcommand{\nb}{\widetilde{\chi}_2^0}

\newcommand{\ncd}{\widetilde{\chi}_{3,4}^0}
\newcommand{\stau}{\widetilde{\tau}}
\newcommand{\mna}{m_{\tilde\chi_1^0}}
\newcommand{\mnb}{m_{\tilde\chi_2^0}}
\newcommand{\mstau}{m_{\tilde{\tau}}}

\newcommand{\mstaua}{m_{\tilde{\tau}_1}}

\newcommand{\mpp}{m_{\pi\pi}}
\newcommand{\mtt}{m_{\tau\tau}}

\newcommand{\pd}{\text{d}}
\newcommand{\tmix}{\theta_{\tilde\tau}}
\newcommand{\ObsMax}{\mathcal{O}_{\text{max}}}
\newcommand{\ObsFrac}{\mathcal{O}_{\nicefrac{1}{10}}}

\newcommand{\EP}{m_{\tau\tau}^{\mathrm{max}}}
\newcommand{\ifb}{$\mathrm{fb}^{-1}$}

\newcommand{\ETM}{\slashed{E}_T}

\newlength{\hoch}

\title{Measuring $\tau$-polarisation in $\nb$ decays at the LHC}

\author{
	Till Nattermann, 
	Klaus Desch,
	Peter Wienemann, 
	Carolin Zendler \\ 
	Physikalisches Institut, University of Bonn, Nu\ss{}allee 12, 53115 Bonn, Germany \\ 
	E-mail: \email{nattermann@physik.uni-bonn.de}, 
	\email{desch@physik.uni-bonn.de},
	\email{wienemann@physik.uni-bonn.de}, 
	\email{zendler@physik.uni-bonn.de}
}

\abstract{We show how the sum of the two average tau polarisations in the decay chain
$\nb \rightarrow \stau_1 \tau \rightarrow \tau \tau \na$ in minimal supersymmetry with conserved $R$-parity can be measured at the LHC.
This is accomplished by exploiting the polarisation dependence of the visible di-tau mass spectrum.
Such a measurement provides information on the couplings of the involved SUSY particles
and allows  a more precise determination of the di-tau mass endpoint. If different tau decay modes can be distinguished,
the polarisation and endpoint measurement can be improved even further.}


\preprint{}

\begin{document}

\section{Introduction}

	If  Supersymmetry (SUSY) is realised in Nature at the TeV scale, the Large Hadron Collider (LHC) has a great 
	potential to copiously produce sparticles. After discovery of new physics, the determination 
	of its properties will be an essential task.
	This objective requires to perform as many measurements as possible to pin down the model. The vast majority 
	of previous studies of SUSY at the LHC focused on methods to extract information on the superpartner masses from LHC
	measurements~\cite{Atlas:CSCnew,CMS:TDR}.
	
	\FIGURE[ht]{
		\epsfig{figure=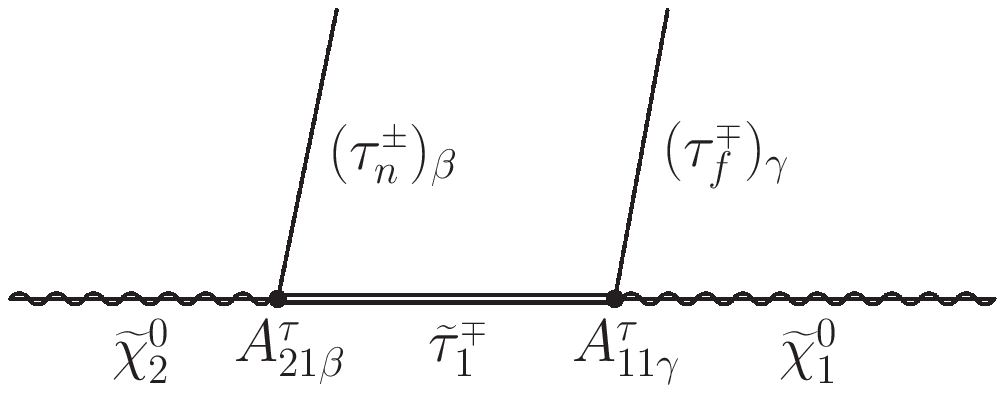,width=0.48\textwidth}
		\caption{signal process, $\beta,\gamma\in\left\{R,L\right\}$}
		\label{Fig:Couplings} 
	}
	
	In this paper we present a technique to obtain information on the average polarisation of $\tau$-leptons originating from
	the decay chain
	\begin{equation}
		\nb \rightarrow \stau_1^{\pm} \tau^{\mp}_n \rightarrow \tau^{\mp}_n \tau^{\pm}_f \na .
	\label{eq:decaychain}
	\end{equation}
	The indices $n$ and $f$ denote \textsl{near} and \textsl{far} to distinguish the two taus according to their mother particle $\nb$ and 
	$\stau_1$, respectively.
	As discussed below, the $\tau$ polarisation provides valuable information about the couplings of $\na$, $\nb$ and $\stau_1$
	and is important for a precise reconstruction of the $\stau_1$ mass. The properties of the $\stau_1$ are of special importance
	for relic density calculations, if the $\stau_1$ is the next-to-lightest supersymmetric particle and is close in mass to the
	lightest SUSY particle (LSP) \cite{Ellis:StauNeutralino}.

	In the following we assume $R$-parity conservation. As a consequence, sparticles are produced pairwise,
	each finally decaying into a stable LSP, which escapes detection, impeding the direct reconstruction of SUSY mass peaks.
	Furthermore, the existence of decay chain~(\ref{eq:decaychain}) 
	implies the mass hierarchy \mbox{$\mnb > \mstaua > \mna$}. We also assume that the considered di-tau final state is not significantly
	contaminated by other SUSY decays (e.~g.~three-body decays of $\nb$, ~\mbox{$\nb \rightarrow \stau_2^{\pm} \tau^{\mp}_n \rightarrow 
	\tau^{\pm}_n \tau^{\mp}_f \na$} or \mbox{$\ncd \rightarrow \stau_1^{\pm} \tau^{\mp}_n \rightarrow \tau^{\pm}_n \tau^{\mp}_f \na$}).

	We exploit the shape of the visible di-tau mass spectrum $\mtt$ as polarisation sensitive observable. Since the energy of the tau-lepton decay
	products in the lab system is	sensitive to its polarisation $P$ \cite{Bullock:TauPhys}, the invariant mass distribution of the visible final 
	state is also affected by the polarisations $P_n$ and $P_f$ of the two taus $\tau_n$ and $\tau_f$ \cite{Zerwas:TauPol,Graesser:ThirdGenSUSY}. The
invariant
	mass distribution of the undecayed taus, $\widehat{m}_{\tau\tau}$, has a triangular shape with a sharp drop-off at a maximum mass
	value
	\begin{equation}
		\widehat{m}_{\tau\tau}^{\text{max}} = \mtt^{\text{max}}=\sqrt{\mnb^2\left(1-\frac{\mstaua^2}{\mnb^2}\right)\left(1-\frac{\mna^2}
		{\mstaua^2}\right)}\, \label{eq:EP}
	\end{equation}
	(see dashed-dotted line in Figure~\ref{FigPionSpectra}). The endpoint $\widehat{m}_{\tau\tau}^{\text{max}}$ of the undecayed di-tau mass
	(including the neutrinos) is equal to the endpoint $\EP$ of the visible di-tau mass, since there is always a certain
	probability that $\widehat{m}_{\tau\tau}$ acquires its maximum value and in both tau decays the complete tau energy is transferred to
	the visible decay products. As the reconstruction of SUSY mass peaks is impossible, the determination of endpoints
	like Eq.~(\ref{eq:EP}) offers a convenient way to obtain SUSY mass information. The considered tau final state is of special
	interest, as it is the only di-lepton final state providing information on $\mstaua$.

Unfortunately, in case of tau-leptons the endpoint measurement is more involved than for stable\footnote{Here {\em stable} means stable on
the scale of the detector size.} leptons, since the visible mass spectrum $\mtt$ slowly fades out and gets very flat at large masses due to the
escaping neutrinos
from tau decays. Previous studies have shown that this problem can be overcome by choosing a suitable endpoint sensitive
observable~\mbox{\cite{Atlas:CSCnew}}, e.~g.~the inflection point of the trailing edge of the mass spectrum. It turned out, that even
for low integrated luminosities, the impact of tau polarisation on this
observable can already exceed its
statistical uncertainty \cite{Atlas:CSCnew}. In this case, polarisation is the largest systematical uncertainty on the $\EP$
measurement
and any
progress on the
$\mstaua$ determination relies on a proper consideration of tau polarisation effects. 

On the other hand, taus are the only leptons providing information on their polarisation. This makes taus a unique probe of new physics giving access
to information which is inaccessible in other measurements.

This paper is structured as follows: In Section~\ref{sec:taupol} the dependence of the polarisation of near and far taus on the underlying SUSY
parameters is summarised. Section~\ref{sec:shapes} discusses the shapes of the visible di-tau mass spectra as function of the tau polarisation for
different tau decay modes. Finally, Section~\ref{sec:study} presents the results of a Monte Carlo study using a parametrised fast simulation of
an LHC detector~\cite{Was:ATLFAST}. It is analysed how tau polarisation information can be extracted from the visible spectra for two different
experimental scenarios: With and without discrimination of the tau decay modes. As an illustration, Section \ref{sec:StauMixMass} discusses the
possible implications
of such a measurement for the stau mixing angle and the stau mass under the assumption, that the $\na$ and $\nb$ properties are already known from
other measurements.

\section{Polarisation of the taus} \label{sec:taupol}

	The polarisation of the considered taus is determined by the coupling constants between $\tau_\beta$, 
	\mbox{$\;\beta\in\left\{R,L\right\}$}, $\stau_1$, and 
	$\widetilde\chi^0_j$. These are given by \mbox{$igA^\tau_{j1\beta}$}, 
	 which are shown in Figure~\ref{Fig:Couplings}. On tree level \mbox{$A^\tau_{j1\beta}$} can be 
	written in terms of the mixing parameters of the neutralinos and the stau \cite{Nojiri:TauPol}:
	\begin{eqnarray}
		A^\tau_{j1L}&=&-\frac{m_\tau}{\sqrt{2}m_W\cos\beta}N^*_{j3}\sin\tmix+\;\frac{1}
		{\sqrt{2}}\left(N^*_{j2}+N^*_{j1}\tan\theta_W\right)\cos\tmix \label{CoupL} \\
		A^\tau_{j1R}&=&-\frac{m_\tau}{\sqrt{2}m_W\cos\beta}N_{j3}\cos\tmix-\;\sqrt{2}N_{j1}\tan\theta_W\sin\tmix\,, 
		\label{CoupR}
	\end{eqnarray}
	where $j=1(2)$ for the coupling involving $\na(\nb)$, $N_{ji}$ are the entries of the neutralino mixing 
	matrix in the $\widetilde B$, $\widetilde W$, $\widetilde H_d$ and $\widetilde H_u$ basis 
	\cite{Zerwas:Neutralino}, \mbox{$\tan{\beta}=\langle H_u\rangle/\langle H_d\rangle$} is the ratio of the two Higgs vacuum expectation values and
	$\tmix$ the stau mixing angle.  Here, only the lighter $\stau_1$ is considered. Within the chosen parametrisation of 
	$\stau$-mixing, \mbox{$\mid\stau_1\rangle=\cos\tmix\mid\stau_L\rangle+\sin\tmix\mid\stau_R\rangle$}, the 
	unmixed case is for \mbox{$\tmix=\pi/2$} resulting in $\stau_1=\stau_R$ and $\stau_2=\stau_L$.

	The first terms of Eqs.~(\ref{CoupL}) and (\ref{CoupR}) are the \textsc{Yukawa} couplings of the 
	$\widetilde H_d$ contributions to the neutralino state, preferring left chiral taus, $\tau_L$, for 
	a $\stau_R$ dominated $\stau_1$ ($\sin\tmix\approx1,\,\cos\tmix\approx0$) and vice versa. This is 
	due to the chirality flipping nature of the \textsc{Yukawa} coupling of the \textsc{Higgs}inos. The 
	$\widetilde H_u$ has no contribution, as it does not couple to the down-type tau. The other terms are the 
	electroweak gauge couplings, which conserve chirality and therefore the coupling 
	to right (left) chiral taus is proportional to the $\stau_{R(L)}$ contribution $\sin\tmix\;(\cos\tmix)$. 
	As the $\widetilde W$-field only couples to left chiral particles, the coupling to right chiral taus has no sensitivity to the wino contribution
	$N_{j2}$. The coupling to the $\widetilde B$ 
	contribution $N_{j1}$ is enhanced by a factor of two due to the double hypercharge carried by right 
	chiral particles compared to left chiral particles.

	With these couplings, the average polarisation is given by
	\begin{eqnarray}
		P=\frac{\left(A^\tau_{j1R}\right)^2-\left(A^\tau_{j1L}\right)^2}{\left(A^\tau_{j1R}\right)^2
		+\left(A^\tau_{j1L}\right)^2}\;, \label{PolTaus}
	\end{eqnarray}
	where $j=1(2)$ has to be chosen for the $\tau_{f(n)}$.
	
	Due to the scalar nature of the $\stau_1$, no angular momentum information can be exchanged between near and far tau. Therefore the realised
	polarisation state of the far tau is independent of that from the near tau in one decay chain. This is leading to the chiralities of the two
	taus being uncorrelated.

\section{Shapes of mass spectra}\label{sec:shapes}

	In the following Section we discuss the impact of tau polarisation on the visible di-tau mass spectrum for different tau decay modes.
	Our line of argument starts from the following assumptions: 1.~The di-tau mass spectrum for undecayed taus has the triangular shape given by phase
	space. 2.~There are no spin correlations between near and far taus.

\subsection{The decay $\tau\rightarrow\pi\nu_\tau$}

	The simplest case is the decay \mbox{$\tau\rightarrow\pi\nu_\tau$}. Due to the fixed chirality of 
	the $\nu_\tau$, as well as momentum and angular momentum conservation, the pion has a preferred emission 
	direction depending on the tau polarisation $P$. The angular distribution of the pion is described by \cite{Bullock:TauPhys}
	\begin{equation}
		\frac{1}{\Gamma_{\tau}}\frac{\text{d}\Gamma_{\pi}}{\text{d}\cos(\vartheta)}=\text{BR}\left(\tau
		\rightarrow\pi\nu_{\tau}\right)\frac{1}{2}\left(1+P\cos\vartheta\right), \label{PionFrag}
	\end{equation}
	where $\vartheta$ is the angle between the pion and tau momentum. The pion 
	coming from right (left) chiral taus is preferentially emitted along (against) the tau momentum, leading to 
	harder (softer) pions. This is reflected in the di-pion spectrum, if both taus decay via $\tau\rightarrow\pi\nu_\tau$. The spectrum
	is given by (see Appendix \ref{Sec:DiPionSpectra})
	\begin{eqnarray}
		N(\mpp)&=&4m\bigg\{\big(P_n\cdot P_f\big)\Big[\ln m\left(\ln m + 4m^2+4\right)+4\left(1-m^2\right)\Big]+ 
		\nonumber \\
		&&\hphantom{4m\bigg\{}+\big(P_n+P_f\big)\Big[m^2-2\ln m-1-\ln^2m\Big]+\ln^2m\bigg\}\;,\qquad \label{EqPiSpectra}
	\end{eqnarray}
	where $m=\mpp/\EP$ is the di-pion mass normalised to the endpoint.
	
	\FIGURE[p]{
		\epsfig{figure=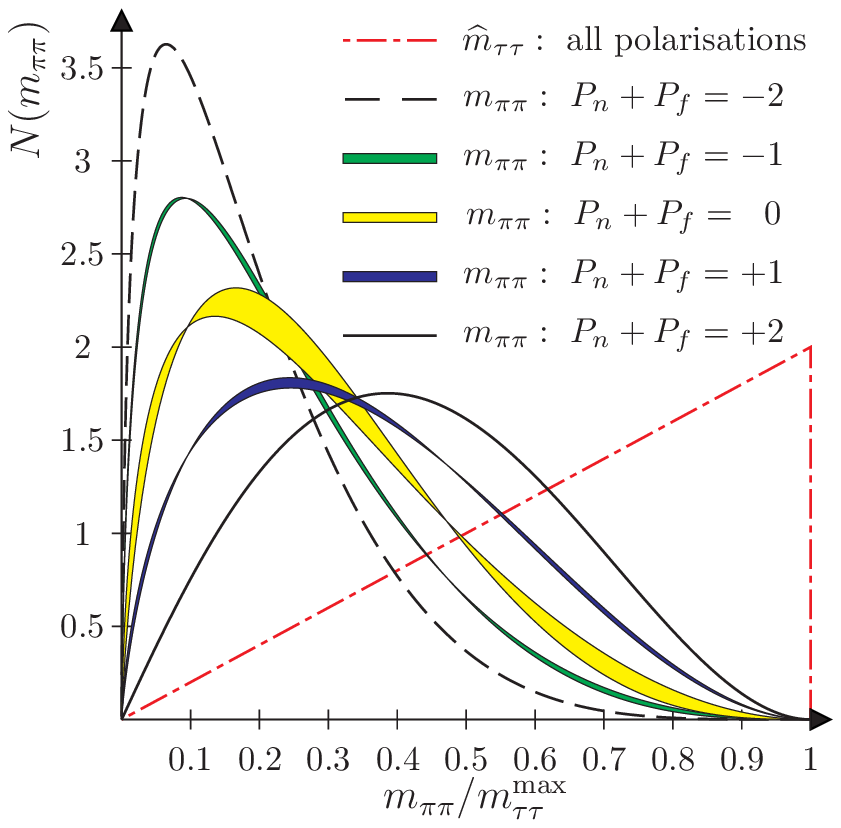,width=0.53\textwidth}
		\caption{$\mtt$ and $\mpp$ distributions}
		\label{FigPionSpectra}
	}

	The polarisations $P_n$ and $P_f$ of the near and far tau enter the spectrum in two different ways: through the sum 
	\mbox{$P_n+P_f$} and the product \mbox{$P_n\cdot P_f$}. Spectra 
	for different sums of the two polarisations are shown in Figure \ref{FigPionSpectra} together with bands indicating the possible 
	variations due to the product terms. For the product terms only those polarisation values are considered which are compatible with the
	respective sum of the polarisations. 
	Obviously, the dominant polarisation effect can be described by the parameter $P_n+P_f$, 
	whereas the product \mbox{$P_nP_f$} has only a subordinate contribution.

	For the cases \mbox{$P_n+P_f=\pm2$}, the product is fully characterised by the sum, whereas for the case of \mbox{$P_n+P_f=0$} the
	variation due to $P_n\cdot P_f$ is most striking as here the individual pairs of $P_n$ and $P_f$ satisfying \mbox{$P_n+P_f=0$} can vary most,
	e.~g.~\mbox{$P_n=P_f=0$} or \mbox{$P_n=\pm1$} and \mbox{$P_f=\mp1$}.

	In principle, a measurement of the $\mpp$-spectrum allows the determination of the endpoint, as well as 
	the sum and the product of the polarisations $P_n$ and $P_f$, although, an extraction of \mbox{$P_nP_f$} is challenging, as the effects due 
	to the product terms in Eq.~\ref{EqPiSpectra} are small compared to the contributions due to \mbox{$P_n+P_f$}.
	If the product could be measured in addition to the sum, both polarisations could be measured individually up to the two-fold ambiguity of
	exchanging \mbox{$P_n\leftrightarrow P_f$}.

\subsection{The decays $\tau\rightarrow\rho\nu_\tau$ and $\tau\rightarrow a_1 \nu_\tau$}

	For the decay of the tau via the $\rho$ and $a_1$ vector meson resonance, the situation is more complicated as the longitudinal 
	and transverse polarisation states of the vector mesons have to be distinguished. For the vector meson case, the fragmentation 
	function is rather complicated \cite{Bullock:TauPhys}, making the calculation of the resulting spectra cumbersome.
	Anyhow, the qualitative behaviour of these decays with respect to the polarisation can be described by 
	the angular distribution of the decay products \cite{Bullock:TauPhys}
	\begin{eqnarray}
		\frac{1}{\Gamma_\tau}\frac{\Gamma_T}{\text{d}\cos\vartheta}&=&\frac{1}{2}\;\text{BR}(\tau\rightarrow
		v\nu_\tau)\frac{2m_v^2}{m_\tau^2+2m_v^2}\left(1-P\cos\vartheta\right) \label{VecTrans} \\
		\frac{1}{\Gamma_\tau}\frac{\Gamma_L}{\text{d}\cos\vartheta}&=&\frac{1}{2}\;\text{BR}(\tau\rightarrow
		v\nu_\tau)\frac{m_\tau^2}{m_\tau^2+2m_v^2}\left(1+P\cos\vartheta\right), \label{VecLong}
	\end{eqnarray}
	where $T\;(L)$ denotes the transverse (longitudinal) vector meson state, $P$ is the tau polarisation, and 
	$\vartheta$ is the angle between the vector meson and the tau momentum. Whereas longitudinally polarised vector mesons 
	have the same angular behaviour as pions, transversely polarised vector mesons behave the opposite way,
	so that the expected net effect depends on the relative branching ratios between transverse and longitudinal 
	states. These are described by the prefactors of Eqs.~(\ref{VecTrans}) and (\ref{VecLong}). With the values 
	$m_\tau=1777$~MeV, $m_{a_1}=1230$~MeV, and $m_\rho=776$~MeV \cite{Yao:PDG}, this leads to approximately the same amount 
	of transverse and longitudinal states for the $a_1$ resonance and more longitudinal ones for the $\rho$. Therefore, 
	the mass spectrum for the decay via $a_1$ is independent of the polarisation, whereas right (left) chiral taus have harder (softer) fragmentation
	for the decay via a $\rho$ meson (similar to $\pi\nu_\tau$-decays).

	Again the spectra can be described by the sum and the product of the polarisations as can be seen from the following argumentation. For a given
	mean polarisation $P$ the probability for a left-chiral tau is \textonehalf$(1-P)$ and that for a right-chiral tau is \textonehalf$(1+P)$.
Therefore the resulting
	spectra can be described by
	\begin{eqnarray} \label{eq:ProdSumVecMes}
		N\left(P_n,P_f\right)&=&\frac{1}{4}(1+P_n)(1+P_f)\big[RR\big]+\frac{1}{4}(1-P_n)(1-P_f)\big[LL\big] + \nonumber \\
		&&+\frac{1}{4}\Big\{(1+P_n)(1-P_f)+(1-P_n)(1+P_f)\Big\}\big[RL\big],
	\end{eqnarray}
	if $\big[RR\big]$, $\big[LL\big]$, and $\big[RL\big]$ are the spectra for two right-chiral taus, two left-chiral taus and the mixed case,
	respectively. Eq.~(\ref{eq:ProdSumVecMes}) leads to the same structure as Eq.~(\ref{EqPiSpectra}), beside the different distributions being
	proportional to the polarisation terms $P_n+P_f$ and $P_n\cdot P_f$. This argument holds in general as long as the there are no spin correlations
	such that the probability for the realised combined polarisation state can be factorised for the two taus.

	\FIGURE[t]{
		\epsfig{figure=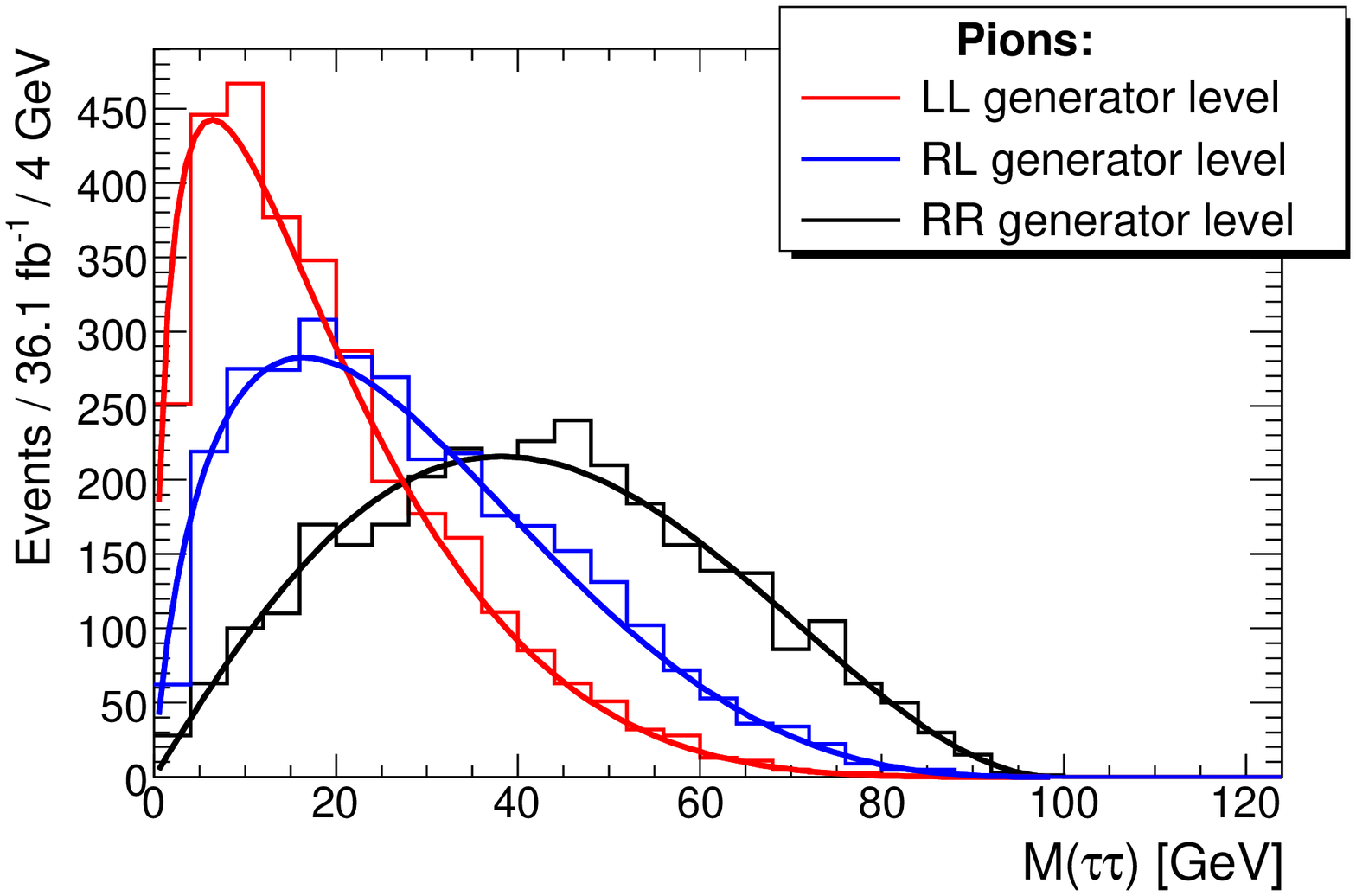,width=0.325\textwidth}
		\epsfig{figure=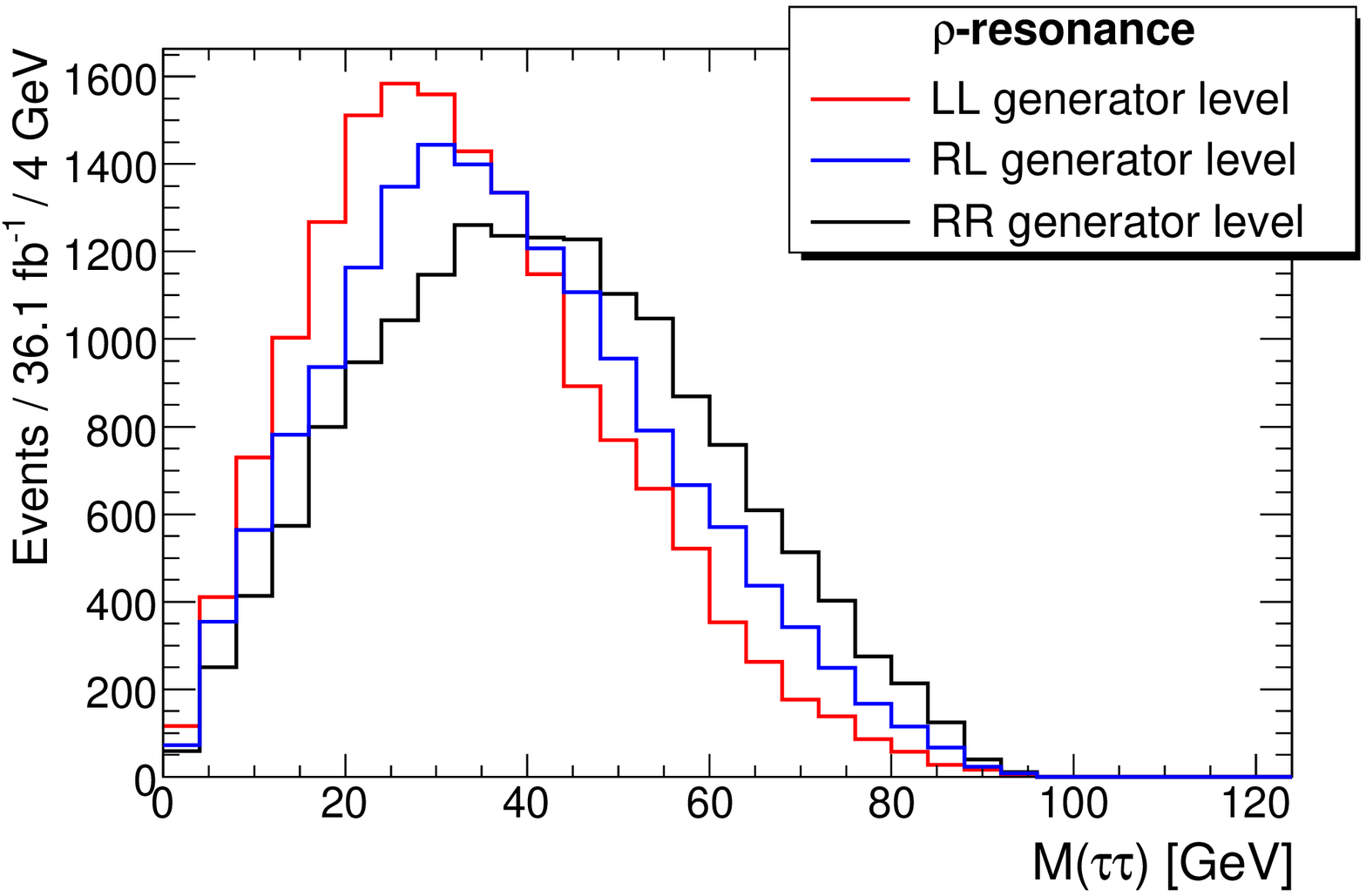,width=0.325\textwidth}
		\epsfig{figure=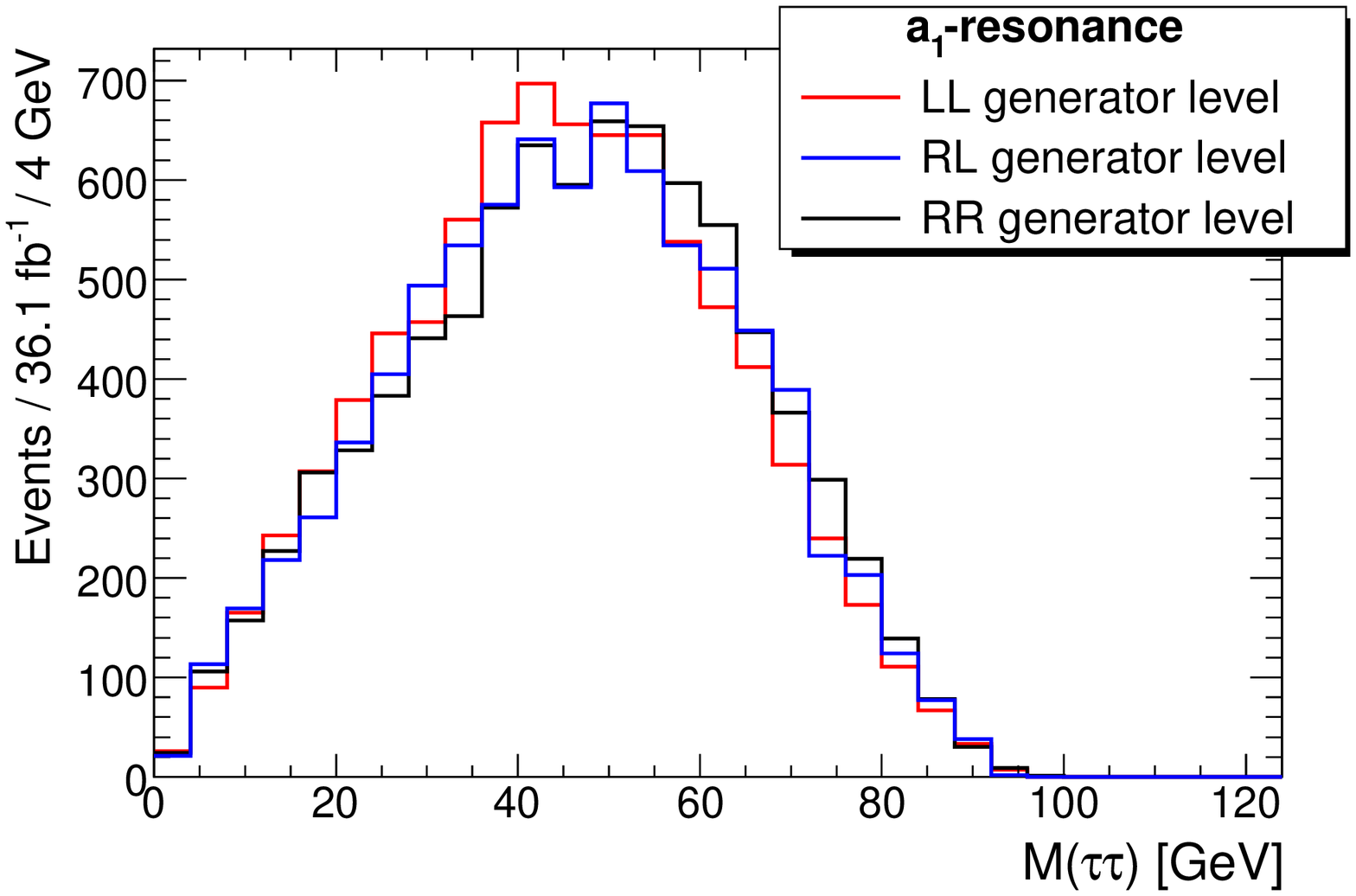,width=0.325\textwidth}
		\caption{Visible mass spectra for different polarisations and different decay modes. In each plot both taus decayed
			in the same channel. The curves are the functions of Eq.~\ref{EqPiSpectra}.}
		\label{FigSpectraDecayMode}
	}
	
	The visible invariant mass of the two taus for different decay modes is shown in Figure~\ref{FigSpectraDecayMode}
	for the cases of \mbox{$P_n=P_f=1$} (RR), \mbox{$P_n=P_f=-1$} (LL) and \mbox{$P_n=\pm1,\;P_f=\mp1$} (RL) as 
	obtained from the generator. The curves are the theoretical shape of Eq.~(\ref{EqPiSpectra}) for the decay into 
	single pions. In each plot both taus decay via the same channel. As expected, the decay over the $a_1$ resonance 
	yields a spectrum being approximately independent of polarisation, whereas the decays into single pions and over the $\rho$ 
	resonance are affected in a similar manner.
	
	In summary, the spectra can be described to a good approximation by the parameters $\EP$ and the sum of the 
	polarisations, \mbox{$P_n+P_f$}, and the impact of the polarisation depends on the decay mode.

\section{Monte Carlo studies} \label{sec:study}

	In the following we present an analysis method which allows to extract $\EP$ and \mbox{$P_n+P_f$} from the visible di-tau mass spectrum. The
	feasibility for such a measurement is illustrated for the mSUGRA bulk region scenario SU3 (\mbox{$M_{\nicefrac{1}{2}}=300$~GeV},
	\mbox{$M_{0}=100$~GeV}, \mbox{$A_0=-300$~GeV}, \mbox{$\tan\beta$=6} and \mbox{$\text{sgn}(\mu)=+1$}, \cite{Atlas:CSCnew}) using an integrated
	luminosity of $36$~\ifb of $pp$ collisions at $\sqrt{s}=14$~TeV. The SUSY sample is generated with Herwig v.6.510
\cite{Corcella:Herwig,Marchesini:Herwig,Moretti:HerwigSUSY}, where the polarisation dependent tau decay is performed by Tauola
\cite{Was:Tauola,Golonka:Tauola}.
	QCD, $t\bar t$, $W+$Jets and $Z+$Jets background are generated with ALPGEN v.2.06 \cite{Mangano:Alpgen}.  To simulate the response of an LHC-type
detector, a parametrised fast detector simulation \cite{Was:ATLFAST} is used as an example.

Due to the simplifications of the fast detector simulation and various idealising assumptions on the performance of the detector and reconstruction
algorithms made in the following, the obtained results should be considered as a demonstration of feasibility rather than a precise evaluation 
of the expected detector performance. A detailed experimental analysis is beyond the scope of this study. A study using full simulation of the ATLAS
detector showed that the decay chain of eq.~\ref{eq:decaychain} can be selected with high purity \cite{Atlas:CSCnew}.

	We extend the technique used in \cite{Atlas:CSCnew}, such that it also includes polarisation effects. To achieve this, we choose
two observables, one of which ideally being maximally sensitive to the endpoint, the other one maximally sensitive to $P_n+P_f$. The relation between
these observables
and the endpoint and the sum of the polarisations is established through a calibration procedure. The calibration is performed using 25 SUSY reference
samples, where every combination of \mbox{$\EP=80,90,100,110,120$}~GeV and \mbox{$P_n+P_f=-2,-1,0,1,2$} 
	is realised. For these samples, only $\mna$, $\mnb$, $\mstau$, $P_n$ and $P_f$ are changed 
	to obtain the corresponding endpoint and sum of the polarisations, whereas all other parameters are 
	fixed to the nominal values of the SU3 benchmark point. 

        For suppression of standard model background,
	typical SUSY cuts (\mbox{$\ETM\ge230$}~GeV, \mbox{$p_{T,1.\text{Jet}}\ge220$}~GeV, \mbox{$p_{T,2.\text{Jet}}\ge50$}~GeV, 
	\mbox{$p_{T,4.\text{Jet}}\ge40$}~GeV) are applied. Due to the large amount of missing transverse energy and the large number 
	of hard jets in SUSY events as well as the two opposite sign tau signature, contributions due to background are 
	negligible \cite{Atlas:CSCnew}.

	To reconstruct the di-tau mass spectrum, the invariant mass of all reconstructed
	oppositely charged 
	taus in one event is built for all events passing the selection cuts. Contributions due to fake 
	taus\footnote{E.~g.~non-tau jets, being identified as a tau-lepton.} and taus not originating from the signal process (combinatorial background),
	can be efficiently 
	suppressed by subtracting the corresponding distribution for same sign taus. This is possible, since 
	the combination of uncorrelated taus has the same probability for equally and oppositely charged 
	taus. Therefore, the background of the opposite sign (OS) distribution due to fakes and wrongly 
	combined taus has the same shape as that of the same sign (SS) taus, since background with correlated OS tau-lepton pairs typically has small
	cross sections (e.~g.~$\widetilde\chi^+_1\widetilde\chi^-_1$ production).

\FIGURE[ht]{
		\epsfig{figure=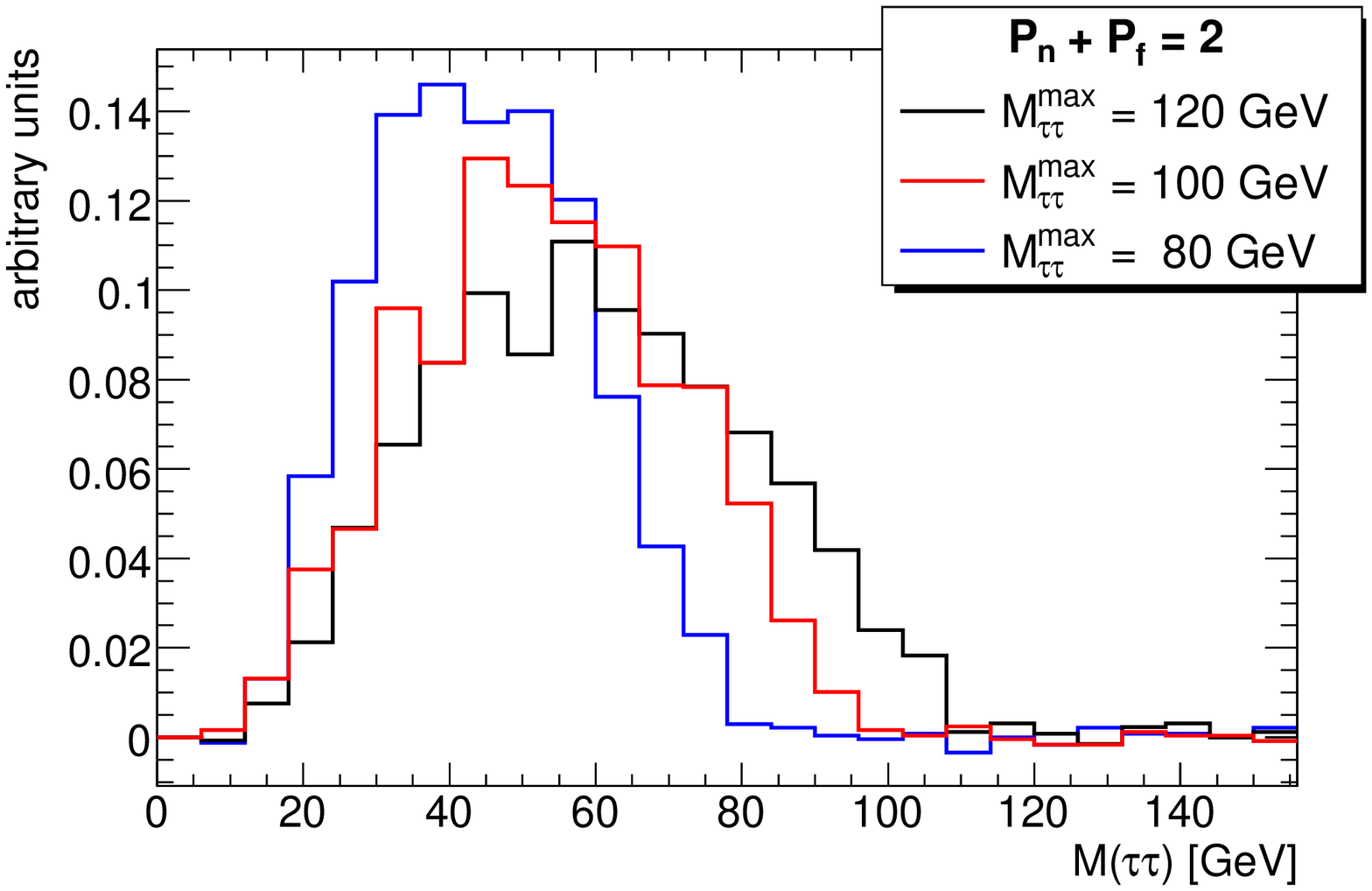,width=0.49\textwidth}
		\epsfig{figure=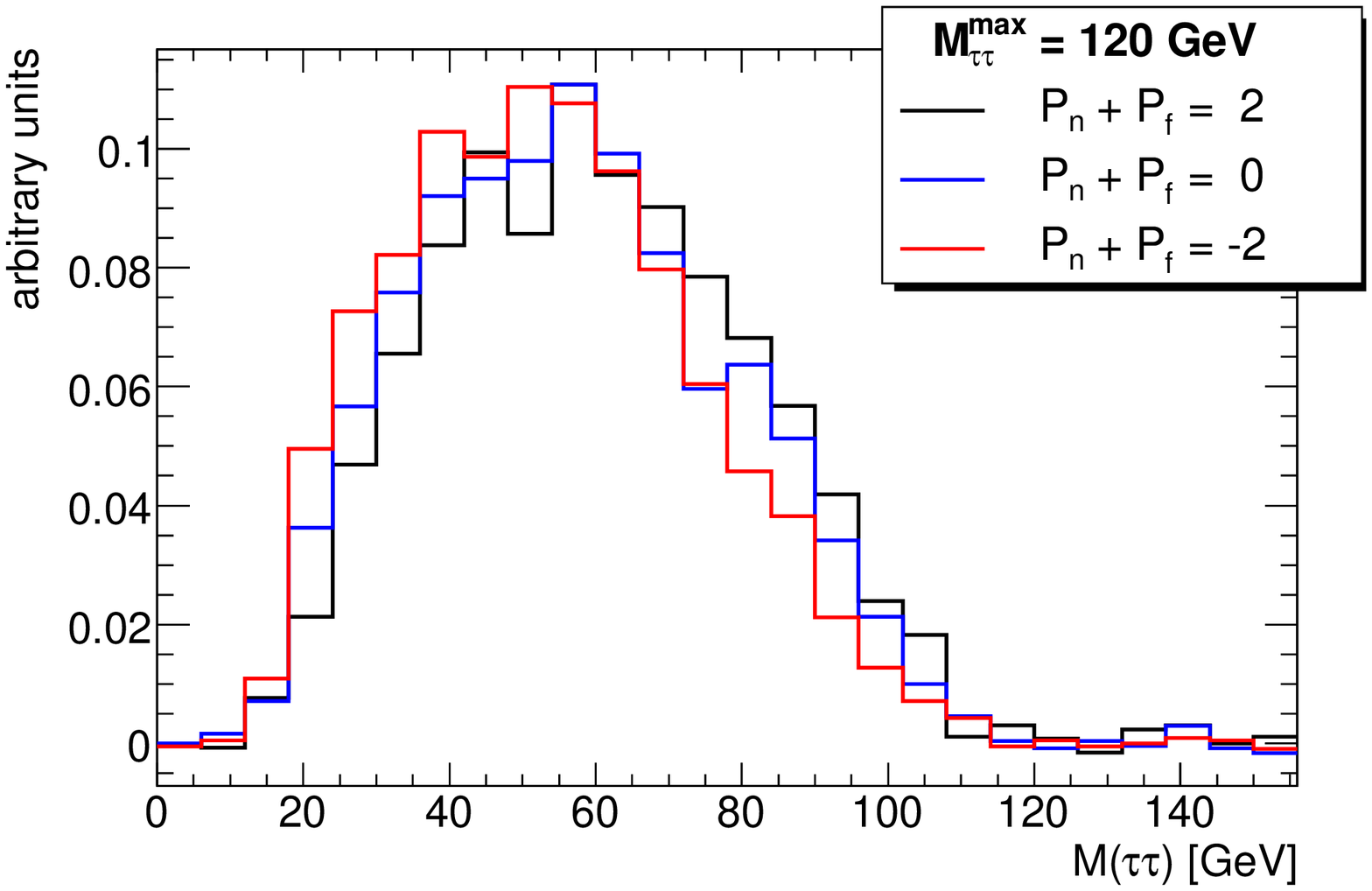,width=0.49\textwidth}
		\caption{Mass and polarisation effects on detector level after reconstruction. The spectra are normalised to unity. Whereas the polarisation
			effects show up most obvious nearby the maximum, different endpoints can be seen mainly in the high mass part of the distributions.
			This will be used to disentangle mass and polarisation effects.}
		\label{FigPolEPsSpectra}
	}
	
\subsection{Without distinguishing tau decay modes}

        Figure \ref{FigPolEPsSpectra} shows the OS-SS mass spectrum for various combinations of 
	endpoint and polarisation, if no distinction between different tau decay modes is made. It is apparent that different polarisation states
	are distinguishable, 
	although the discrimination power is reduced compared to the di-pion spectra of 
	Figure~\ref{FigPionSpectra}. This is due to the fact, that the decay $\tau\rightarrow\pi\nu_\tau$ 
	is affected most by polarisation effects. Moreover some discrimination power in the low mass part of the spectra is lost, as the tau 
	reconstruction is less efficient for low $p_T$-taus.

	Figure \ref{FigPolEPsSpectra} also shows, that SUSY masses and
	tau polarisation have different impact on the spectrum. Whereas the polarisation does not affect
	the endpoint and mostly shifts the position of the maximum, different
	SUSY masses stretch the whole spectra and show up most strikingly in the high mass part of
	the distribution. This will be used to disentangle polarisation and mass effects.

	For all 25 generated reference spectra, a fit with a Gaussian\footnote{Among many tested functions, a Gaussian turned out to be the best
compromise between shape agreement and mathematical simplicity.} is performed by \texttt{Minuit} 
	\cite{James:Minuit} and from this the position of the maximum, $\ObsMax$, as well as the 
	position, where the fit decreased to $\nicefrac{1}{10}$ of its maximum value at the high mass part, 
	$\ObsFrac$, is calculated. Since the low mass part of the spectrum 
	is strongly affected by the effects of the tau reconstruction, which does not contain 
	any information about SUSY, the lower fit range is chosen to be the first bin, exceeding 
	half of the maximal bin content of the spectra. This guarantees a common prescription  
	for fitting the spectra, which is necessary in order to allow better comparability. However, due to statistical 
	fluctuations of the bin content, the chosen range can vary, which needs to be 
	considered in the error calculation.

	\FIGURE[t]{
		\epsfig{figure=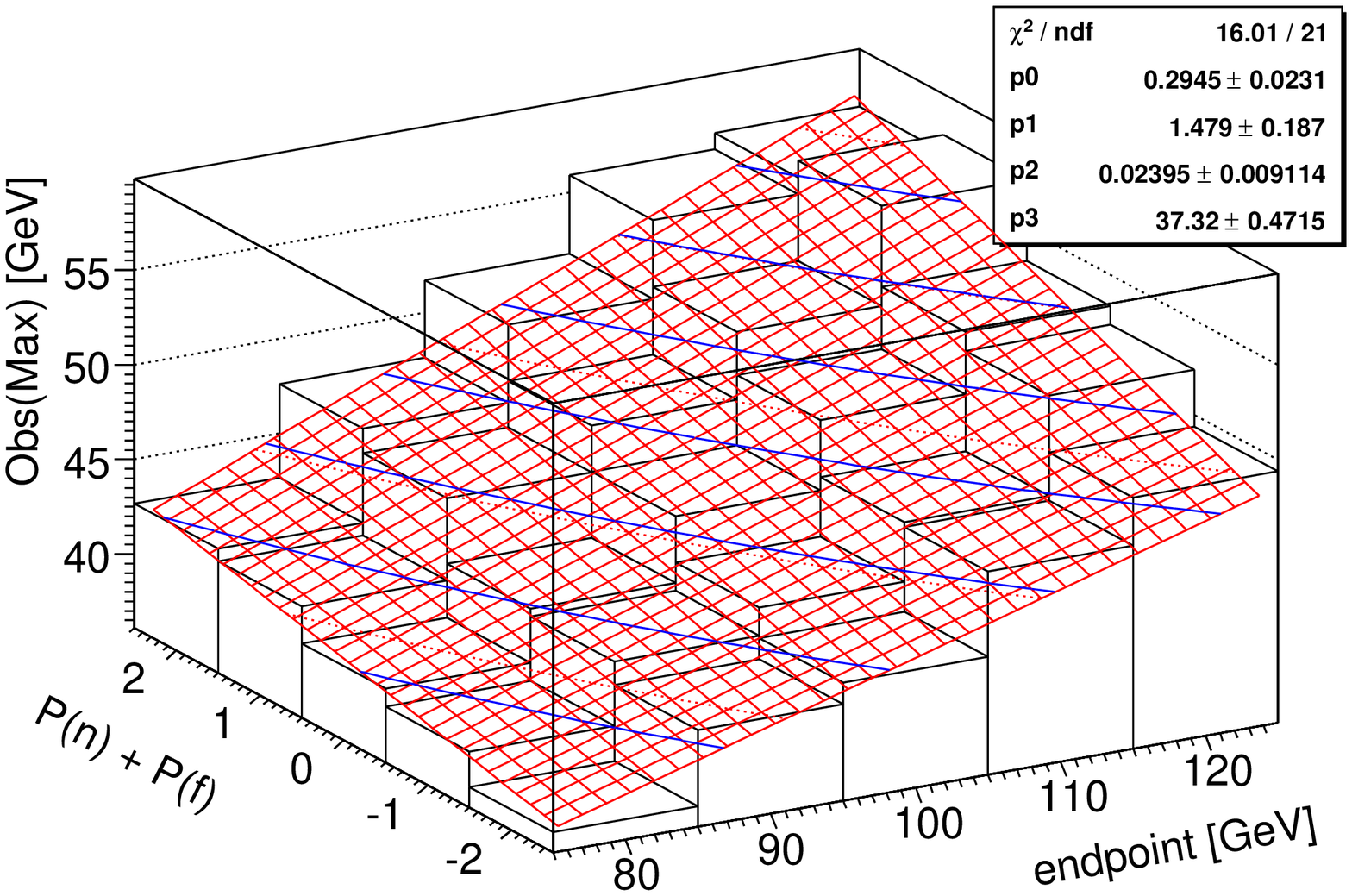,width=0.49\textwidth}
		\epsfig{figure=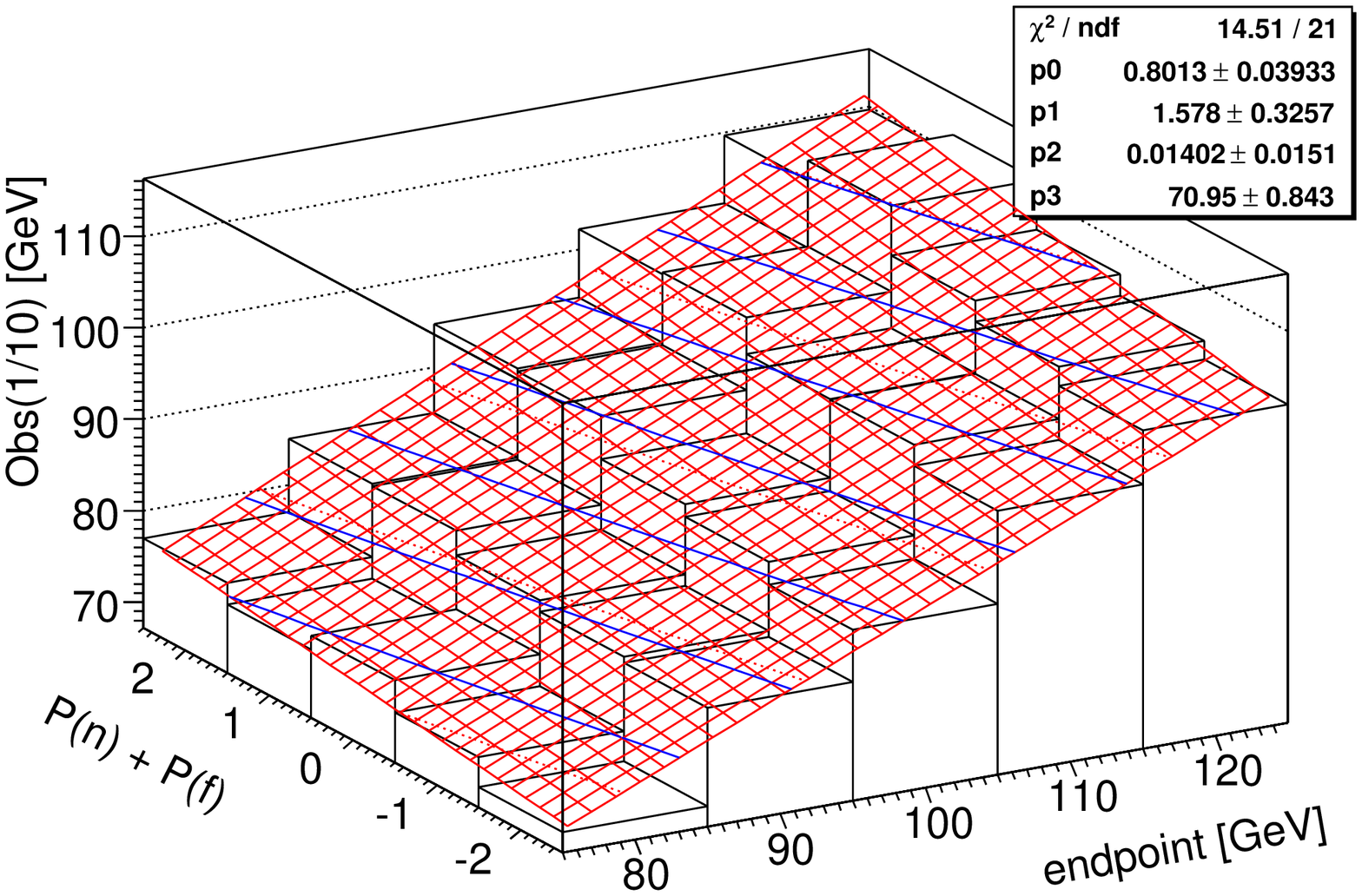,width=0.49\textwidth}
		\caption{The observables $\ObsMax$ (left) and $\ObsFrac$ (right) in the parameter plane of $\EP$ and $P_n+P_f$. The two-dimensional histograms
show the expected value of the observables for each simulated reference point (endpoint, $P_n+P_f$), the red meshes represent the fitted function
of eq. \ref{FitFunct} and the blue lines correspond to equal values of the observables.}
		\label{FigObsMaps}
	}

	The 25 generated reference samples represent a 5$\times$5 grid in the parameter plane spanned by $\EP$ and $P_n+P_f$.
	For each grid point the chosen observables $\ObsMax$ and $\ObsFrac$ and their errors can be calculated from the fitted Gaussian parameters. Figure 
\ref{FigObsMaps} shows the obtained values for the two observables $\ObsMax$ (left plot) and $\ObsFrac$ (right plot) plotted against the
endpoint, $\EP$,  and $P_n+P_f$.

	To obtain a continuous mapping from $\EP$ and $P_n+P_f$ to the observables $\ObsMax$ and $\ObsFrac$, a two
dimensional fit of the function 
	\begin{eqnarray}
		\mathcal{O}^{\text{fit}}\left(P_n+P_f,\mtt^{\text{max}}\right)
		&=&p_0\big(\mtt^{\text{max}}-80\;\text{GeV}\big)+p_1\left(P_n+P_f+2\right)+  \nonumber \\ 
		&&+p_2\big(\mtt^{\text{max}}-80\;\text{GeV}\big)\big(P_n+P_f+2\big)+p_3\,,\label{FitFunct} 
	\end{eqnarray}
	is performed to the grid points for each observable. The fitted functions for the two observables are also included in Figure \ref{FigObsMaps}
(shown in red).
	Both functions increase for larger endpoints, since the whole 
	spectrum gets stretched for shifted endpoints. The dependence on $P_n+P_f$ is quite different:
	Whereas the polarisation has a sizable influence on the position of the maximum, $\ObsFrac$, which is close to the polarisation independent
endpoint, is hardly affected (compare Figure~\ref{FigPolEPsSpectra}). This  difference is reflected in diverse behaviour of the blue contour
lines in left and right plot in Figure~\ref{FigObsMaps}.


	The measured value of one of the observables, $\ObsMax$ and $\ObsFrac$, can then be transferred to corresponding $\EP$ and \mbox{$P_n+P_f$}
	values using the calibration. Each of the two measured observables defines a contour line in Figure~\ref{FigObsMaps}, which provides a set of
	$\EP$ and $P_n+P_f$ values, being compatible with the measurement. The intersection point of these two contour lines finally yields both, the
	endpoint position and the sum of the two polarisations. It is crucial for this method, that the two observables 
	have different dependencies on the endpoint and the polarisation, giving two intersecting contour 
	lines.

	To test the described method, the nominal SU3 spectrum is generated in order to determine 
	the endpoint and the sum of the two polarisations. The two observables are extracted as described above. Statistical uncertainties are
	estimated by \texttt{Minuit}. Systematic uncertainties due to the variation 
	of the fit range are determined by calculating the observables for different choices of the fit range.
	Both contributions are added quadratically, neglecting possible correlations.

	To find the combination of $\EP$ and $P_n+P_f$, fitting best to the measured observables, 
	$\ObsMax^{\text{meas.}}$ and $\ObsFrac^{\text{meas.}}$, the function
	\begin{eqnarray}\label{Chi2Fct}
		\chi^2\left(\mtt^{\text{max}},P_n+P_f\right)=
		\left[\vec{\mathcal{O}}^{\text{fit}}-\vec{\mathcal{O}}^{\text{meas.}}\right]^T
		\text{Cov}^{-1}\left(\vec{\mathcal{O}}\right)
		\left[\vec{\mathcal{O}}^{\text{fit}}-\vec{\mathcal{O}}^{\text{meas.}}\right]
	\end{eqnarray}
	is minimised, where \mbox{$\vec{\mathcal{O}}^{\text{fit}}=\big(\ObsMax^{\text{fit}},\;
	\ObsFrac^{\text{fit}}\big)$} are the expected observables values for given $\EP$ and $P_n+P_f$,  as
	obtained from Eq.~(\ref{FitFunct}) and $\text{Cov}\big(\vec{\mathcal{O}}\big)$ is the corresponding
	covariance matrix. This matrix can be calculated from the covariance 
	matrix of the parameters of the fitted Gaussian. This method ensures the correct treatment of the correlation between $\ObsMax$ and 
	$\ObsFrac$. The uncertainties on $\EP$ and $P_n+P_f$ are given by the
	contours, satisfying $\Delta\chi^2=\chi^2_{\text{min}}+a(\text{CL})$, where $a(\text{CL})$ depends on the desired confidence level.

	\FIGURE[t]{
		\epsfig{file=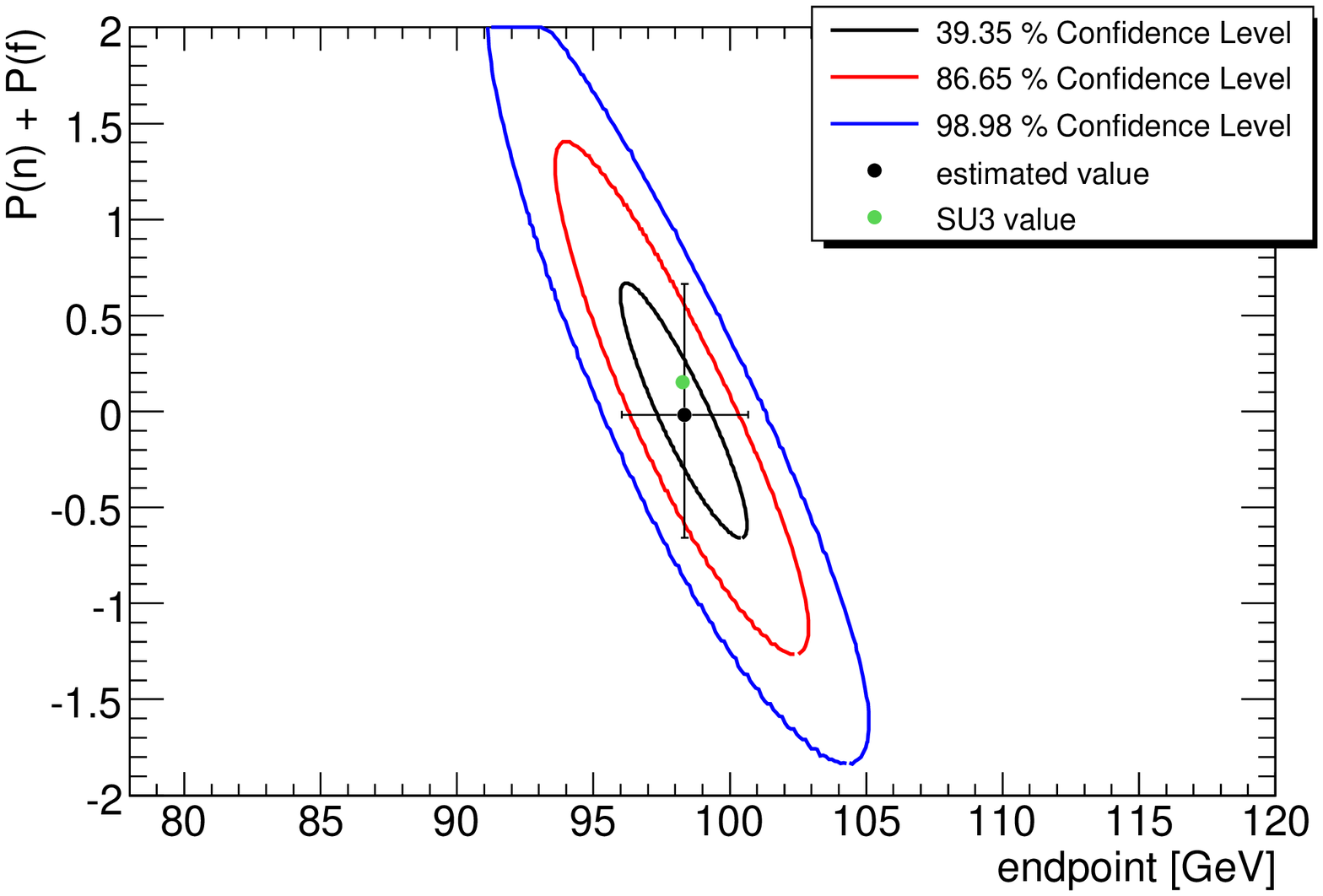,width=0.49\textwidth}
		\epsfig{file=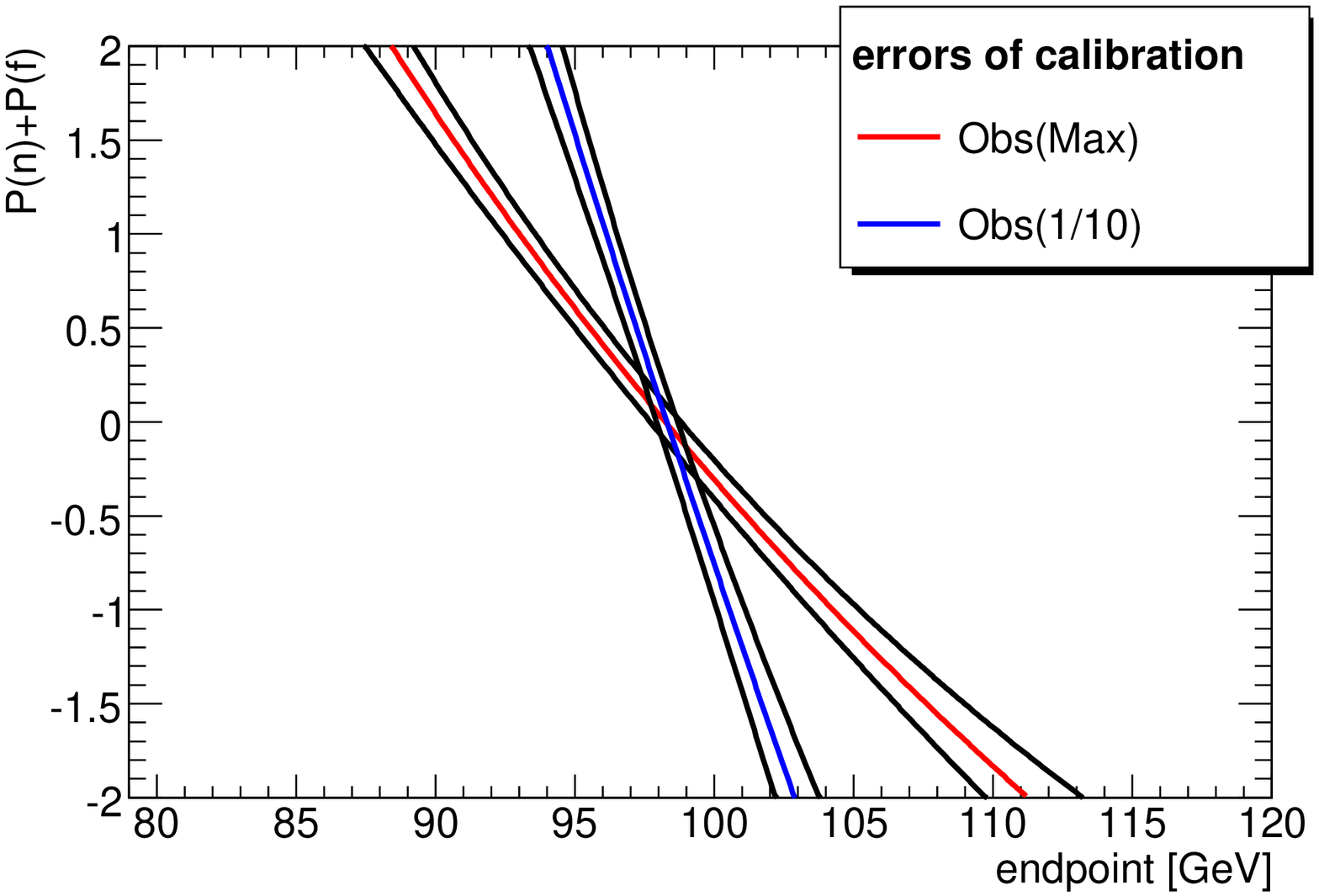,width=0.49\textwidth}
		\caption{Determined values (left) and additional contributions due to uncertainties in the calibration (right). No discrimination between tau
			decay mode is made.}
		\label{FigResult}
	}

	The result can be seen in the left plot of Figure~\ref{FigResult}, where contours for different
 	confidence levels together with the true value of the investigated SU3 point are shown. The 
	determination of the endpoint is rather precise, whereas the polarisation is less constrained by the measurement.
	The tilt of the contours reflects the anti-correlation of the endpoint 
	and the sum of the polarisations. As both observables increase with growing $\EP$ or $P_n+P_f$, 
	a larger $\EP$ is more compatible with a smaller $P_n+P_f$ and vice versa for a fixed observable.

	In the above analysis uncertainties on the parameters $p_i$ in Eq.~\ref{FitFunct} are not considered.  Due to the limited statistics 
	in the calibration samples, the estimated values for $p_i$ in Eq.~\ref{FitFunct}
	have errors, involving an additional uncertainty, which may be reduced in principle. Their contributions to the uncertainty can be seen in the
	right plot of Figure \ref{FigResult}. Here the central value lines obtained from the measured values of $\ObsMax$ 
	and $\ObsFrac$ are shown with corresponding uncertainty bands. These bands are determined by solving
	\begin{equation}
	     \mathcal{O}^{\text{fit}}
	(\EP,P_n+P_f)\stackrel{!}{=}\mathcal{O}^{\text{meas.}}
	\end{equation}
	for $P_n+P_f$ and calculating the derivatives
	\begin{eqnarray}	
		\sigma_y(x)=\;\sqrt{\sum_{i=0}^3\sum_{j=0}^3\frac{\partial y}{\partial p_i}
		\frac{\partial y}{\partial p_j}\text{cov}(p_i,p_j)}\,,
	\end{eqnarray}
	where $y=P_n+P_f$ and $x=\EP$. The effects are small compared to the other uncertainties
	and are neglected. The intersection of the lines is the central value
	of the measurement, since here \mbox{$\chi^2=0$}.

\subsection{With distinguishing tau decay modes}

	\FIGURE[p]{
		\epsfig{figure=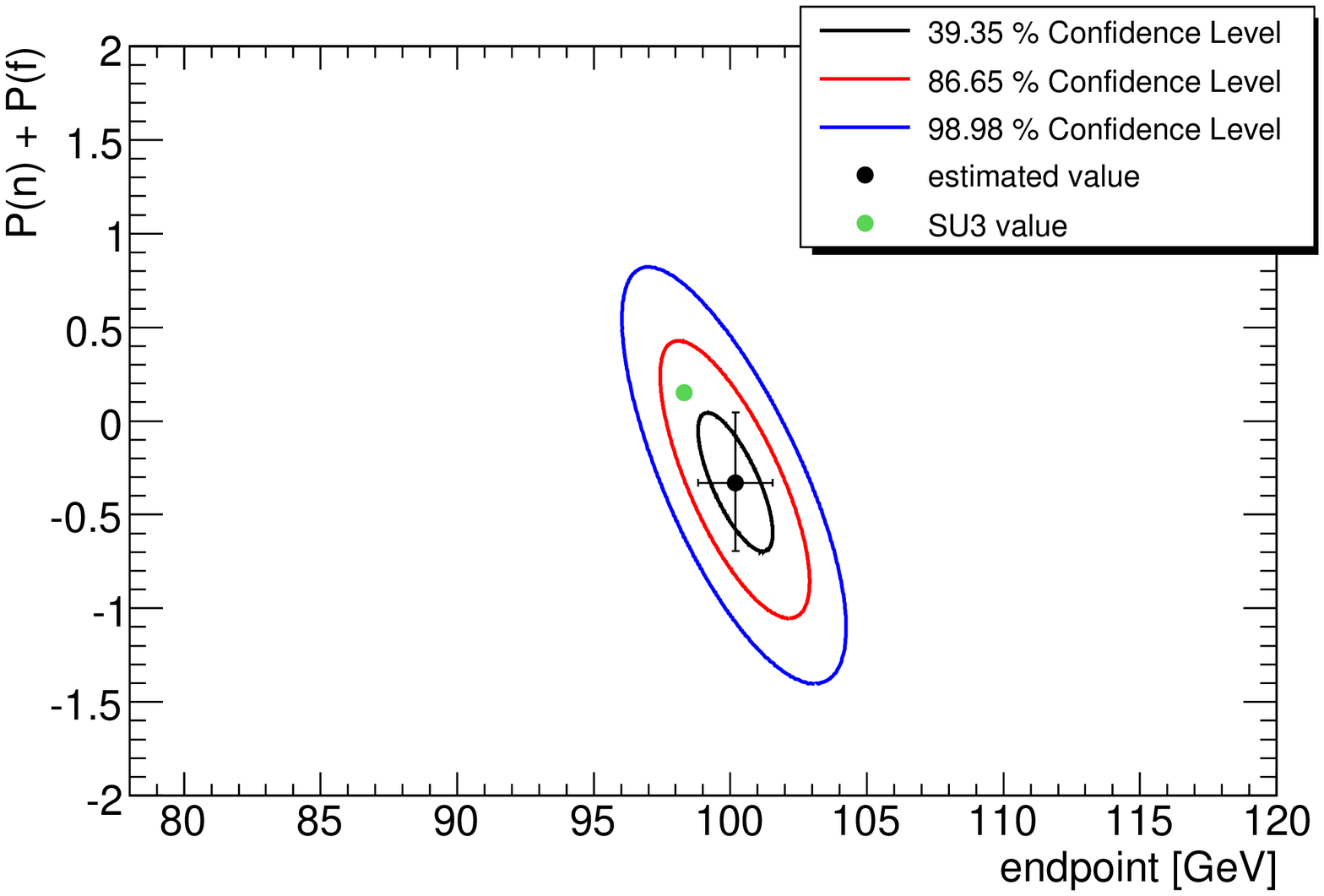,width=0.52\textwidth}
		\caption{Result with idealised tau decay mode discrimination}
		\label{FigResultTauMode}
	}
	
	So far, the shown spectra include all tau decay modes. As shown in the Section \ref{sec:shapes}, 
	the polarisation effects depend on the decay mode. The decay via the $a_1$ vector meson e.~g.~has no 
	sensitivity on the polarisation and therefore the most precise determination of the endpoint 
	can be achieved with events, where both taus decay via the $a_1$ resonance. In contrast to that, the decay 
	$\tau\rightarrow\pi\nu_\tau$ is affected most and is therefore a perfect polarisation analyser. 
   Exploiting these differences between decay modes can improve the precision of the combined endpoint and polarisation measurement.
   In the following we investigate how much improvement can be gained with a more sophisticated analysis taking decay mode
	specific differences into account. For this study we assume an idealised diagonal efficiency matrix for tau reconstruction:
	\begin{equation}\label{eq:EffMatrix}
	   \left( \begin{matrix} N_{\tau\rightarrow X_1} \\ \vdots \\  N_{\tau\rightarrow X_n} \end{matrix}\right)_{\text{reco.}}  =\left(\begin{matrix}
\epsilon &   & 0 \\   & \ddots &  \\ 0 &   & \epsilon \end{matrix} \right)   \left( \begin{matrix} N_{\tau\rightarrow X_1} \\
\vdots \\ N_{\tau\rightarrow X_n} \end{matrix}\right)_{\text{truth}} + N_{\text{Fakes}}\left(\begin{matrix} \text{BR}(\tau\rightarrow X_1) \\ \vdots
\\ \text{BR}(\tau\rightarrow X_n) \end{matrix} \right),
	\end{equation}
	where $\epsilon=\epsilon(p_T,\eta)$ is the reconstruction efficiency averaged over all tau decay modes and $N_{\tau\rightarrow X_i}$ are the
	number of taus decaying to $X_i$. $N_{\text{Fakes}}$ is the number of fake tau-leptons from the parametrised tau-reconstruction algorithm. Fake
taus are randomly  assigned to a certain decay mode according to the respective branching ratio. The mean
reconstruction efficiency for hadronically decaying taus with $0\le\left|\eta\right|\le2.5$ and
$10\;\text{GeV}\le p_T^\text{vis.}\le100\;\text{GeV}$ is 0.33.

Now we analyse the di-tau mass spectrum separately for each tau decay category.
	In order to have reasonable statistics in each spectrum, splitting up all reconstructed 
	di-tau-events into too many distributions is not viable. Hence, only two types
	of decays are differentiated: decays into single $\pi$, single $K$ and $\rho$, 
	which are affected by polarisation, and decays which are 
	not affected (i.~e.~decays via $a_1$ and all remaining decays). In this case one obtains three different cases: the fragmentation of both taus
 is polarisation independent, the fragmentation of both taus is polarisation dependent and the 
	mixed case. Since for each of the three spectra $\ObsMax$ and $\ObsFrac$ 
	are measured, we end up with six observables in this case, each providing a contour line in
	$(\EP$,$P_n+P_f)$ space. The observables are determined in the same manner as for the case without 
	distinguishing the tau decay modes.

	The result can be seen in Figure~\ref{FigResultTauMode}, showing, that the measurement can be significantly
	improved provided that information about the decay mode is available. Both the endpoint and the polarisation measurement are affected. For these
	results the $\chi^2$ function of Eq.~\ref{Chi2Fct} 
	has been minimised using the six observables.

\subsection{Interpretation in terms of $\tmix$ and $\mstau$} \label{sec:StauMixMass}

	The measured parameters $\EP$ and \mbox{$P_n+P_f$} depend on many fundamental parameters of the underlying 
	SUSY model (Eq.~\ref{eq:EP} and \ref{CoupL}--\ref{PolTaus}). Therefore, these quantities cannot be 
	extracted in a direct way from the described measurement. Under the assumption, that the neutralino 
	sector (masses and mixing) is already known from other measurements, the sum of the polarisations is only a 
	function of the stau mixing angle and the endpoint is completely specified by the stau mass. In this 
	case, the probability density \mbox{$f\big(\EP,(P_n+P_f)\big)$} of Figure~\ref{FigResult} can 
	be translated into a probability density \mbox{$g\big(\mstau,\tmix\big)$} via
	\begin{eqnarray}
		g(\mstau,\tmix)=f\Big(\mtt^{\text{max}}\left(\mstau\right),\big[P_n+P_f\big]\left(\tmix\right)\Big)
		\left|\frac{\partial\mtt^{\text{max}}}{\partial \mstau}\cdot\frac{\partial \big[P_n+P_f\big]}
		{\partial \tmix}\right|. \label{WDMixM}
	\end{eqnarray}
   For that, the uncertainty contours of Figure \ref{FigResult} are approximated by ellipses making \mbox{$f$} a 
	2-dimensional Gaussian probability density.

	\FIGURE[t]{
		\begin{minipage}{0.49\textwidth}
		\epsfig{file=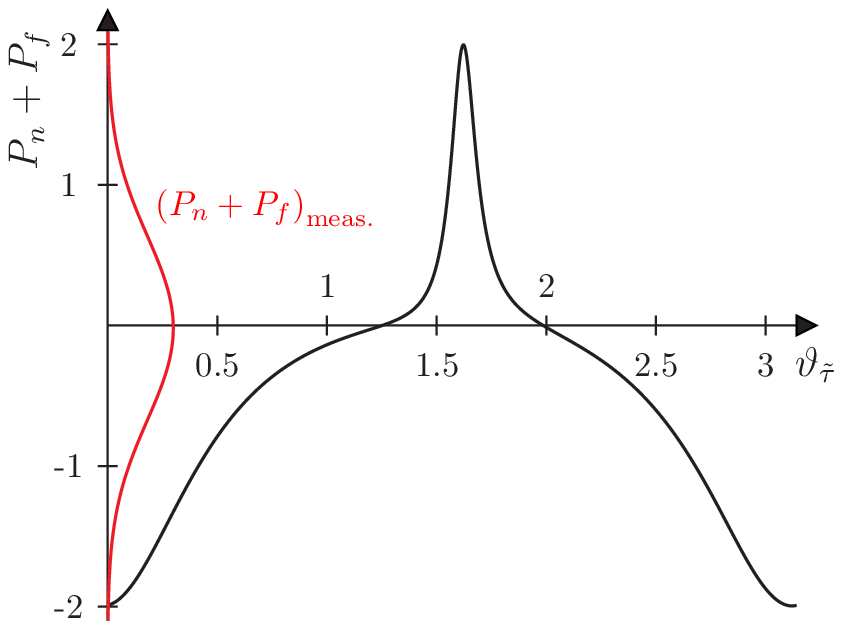,width=\textwidth} \vspace*{0cm}
		\end{minipage}
		\begin{minipage}{0.49\textwidth}
		\epsfig{file=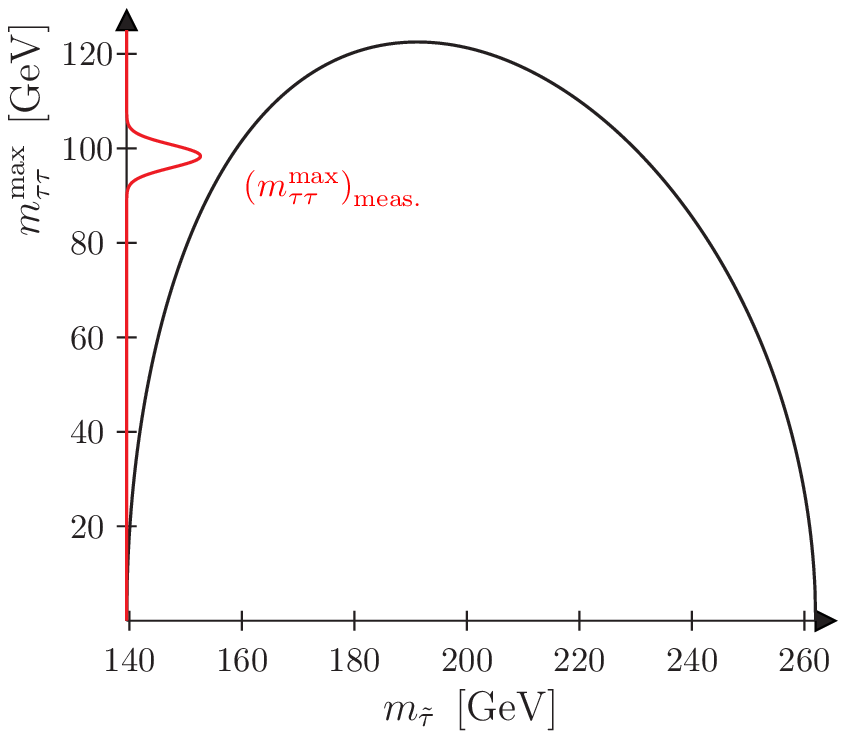,width=\textwidth}
		\end{minipage}
		\caption{The functions $(P_n+P_f)(\tmix)$ and $\EP(\mstaua)$ with the measured values (red).}
		\label{FigPolEP}
	}

For given neutralino masses the maximal possible endpoint of Eq.~(\ref{eq:EP}) is the difference of $\mnb$ and $\mna$. Assuming the nominal
neutralino masses for the SU3 scenario, this upper limit leads to the additional requirement $\EP<101$~GeV, which is more constraining than
our measurement (see Figure~\ref{FigResult}). As this extra restriction is a special feature of the chosen benchmark point, we choose a different
neutralino sector for our following analysis, where the	kinematic limit on $\EP$ is less constraining than the measurement. An example is
the neutralino sector of the SU1 point (\mbox{$M_{\nicefrac{1}{2}}=350$~GeV},
	\mbox{$M_{0}=70$~GeV}, \mbox{$A_0=0$~GeV}, \mbox{$\tan\beta$=10} and \mbox{$\text{sgn}(\mu)=+1$}, \cite{Atlas:CSCnew}). For this scenario the
functions $\EP(\mstau)$ and $(P_n+P_f)(\tmix)$ are displayed in Figure \ref{FigPolEP} together with 
	the expected precision of their measurement, shown as a Gaussian indicating their uncertainties. The
	stau mass must be in the range $139.5\;\text{GeV}=\mna\le\mstaua\le\mnb=262.9\;\text{GeV}$ to allow the subsequent 
	two-body decays, whereas in principle all mixing angles $\tmix$ are possible.

	\FIGURE[ht]{
		\epsfig{figure=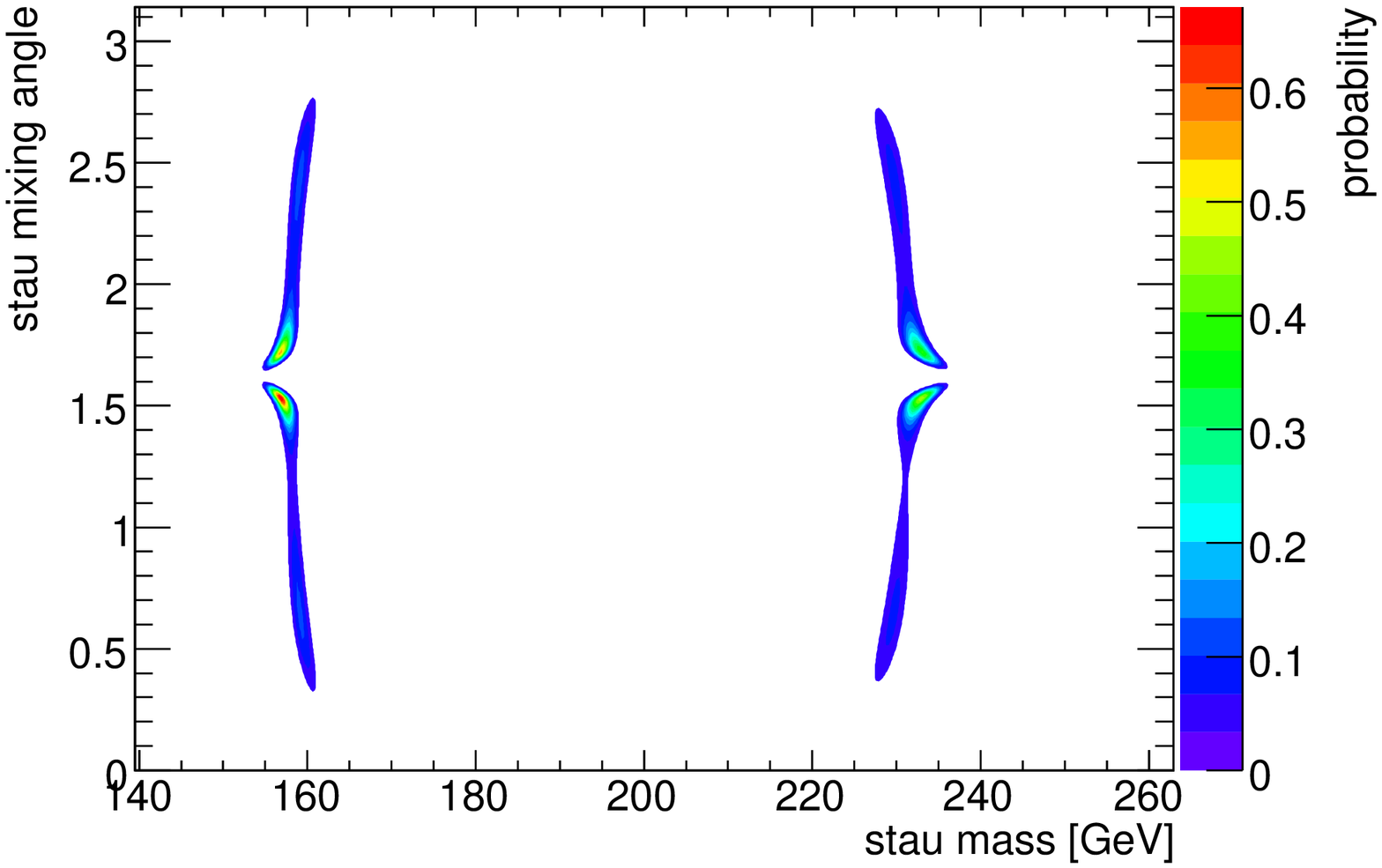,width=0.49\textwidth}
		\caption{$\tmix$ and $\mstau$ determination}
		\label{FigStauMixStauMass}
	}

	The transformed probability density can be seen in Figure~\ref{FigStauMixStauMass}. Since $\EP$
	($P_n+P_f$) is consistent with two values of $\mstau$ ($\tmix$), four regions 
	are possible. The small preferred regions near $\tmix\approx1.5,1.7$ are due to the fact, that 
	here, a small variation on $\tmix$ causes a large variation in $P_n+P_f$ and therefore a large 
	fraction of the possible values of $P_n+P_f>0$ are projected on a small region of $\tmix$, making 
	this values very likely. The long tail towards higher (lower) values is due to the values $P_n+P_f<0$, 
	which are distributed more equably on the $\tmix$-axis (compare left plot of Figure~\ref{FigPolEP}).
	
	The two possible $\mstaua$ are those being consistent with the measured endpoint (compare right plot of Figure 
	\ref{FigPolEP}). The correlation of the parameters $\EP$ and $P_n+P_f$ can be seen in the bending
	of the four regions. 

\section{Conclusions}

	The polarisation dependence of the visible di-tau invariant mass spectrum from the decay \mbox{$\nb\rightarrow\stau_1\tau\rightarrow\na\tau\tau$}
is -- to a good
approximation -- parametrised by the sum $P_n+P_f$ of the near and far tau polarisation. It can be measured by exploiting shape variations for
different polarisation states. Our method used two $P_n+P_f$ and $\EP$ sensitive observables to characterise the spectrum shape and establishes the
relationship between the measured observables and $P_n+P_f$ and $\EP$ through a calibration procedure. From the polarisation measurement,
information on the SUSY couplings can be derived, which might be valuable additional input for global SUSY parameter fits
\cite{Bechtle:fittino,Lafaye:SFitter,Flaecher:GFitter}.

The precision on $\EP$ as
obtained from a combined fit with $P_n+P_f$ can be improved with respect to an endpoint measurement neglecting polarisation effects.
The results can even be improved provided that different tau decay modes can be distinguished sufficiently well.

\acknowledgments

The authors are grateful to Peter Zerwas for fruitful discussions on theoretical aspects of this analysis. We also would like to thank Borut Kersevan
for help on how to modify tau polarisation in the Monte Carlo event generator.

We thank members of the ATLAS Collaboration for helpful discussions. We have made use of the
ATLAS physics analysis framework and tools which are the result of collaboration-wide efforts.

This work was supported by the German Ministry of Science and Education
(BMBF) under contract no. 05 HA6PDA.

\begin{appendix}

\section{Pion invariant mass distribution}\label{Sec:DiPionSpectra}

	The invariant mass of the two undecayed taus can be written as
	\begin{eqnarray}
		\widehat{m}_{\tau\tau}^2&=&\frac{1}{2}\big(1-\cos\vartheta\big)\mnb^2\left(1-\frac{\mstau^2}
		{\mnb^2}\right)\left(1-\frac{\mna^2}{\mstau^2}\right).\label{InvTauTau2}
	\end{eqnarray}
	where $\vartheta$ is the angle between the momenta of the $\tau_f$ and the $\stau_1$ 
	in the rest frame of the $\stau_1$. As the $\stau_1$ is a scalar particle, the 
	probability density for $\widehat{m}_{\tau\tau}$ a linearly increasing function.

	Within the relativistic limit, \mbox{$\beta_\tau=1$}, the probability $F(x)$ 
	of the fractional energy of the tau carried by the pion, \mbox{$x=E_\pi/E_\tau$}, is 
	described by the fragmentation function \cite{Bullock:TauPhys}
	\begin{eqnarray}
		F_i(x_i)=\left[1+ P_i(2x_i-1)\right], \label{EqSinglePionFrag}
	\end{eqnarray}
	where \mbox{$i\in\left\{n,f\right\}$} and $P_{n,f}$ is the polarisation in terms of 
	chirality. To obtain the resulting spectra of the two pions, the 
	probability density $H(z)$ for 
	\begin{eqnarray}
		z=\frac{\mpp^2}{\widehat{m}_{\tau\tau}^2}=\frac{\left(p_{\pi_n}+p_{\pi_f}\right)^2}
		{\left(p_{\tau_n}+p_{\tau_f}\right)^2}\approx\frac{E_{\pi_n}E_{\pi_f}
		\left(1-\cos\varphi\right)}{E_{\tau_n}E_{\tau_f}\left(1-\cos\vartheta\right)}
		\approx\frac{E_{\pi_n}E_{\pi_f}}{E_{\tau_n}E_{ \tau_f}}=x_n\cdot x_f
	\end{eqnarray}
	is needed, where $\varphi$ $(\vartheta)$ is the angle between the pions (taus). The 
	fourth equality only holds in the collinear limit where $\varphi=\vartheta$. In the 
	signal decay chain, the particle "between" the two taus is the scalar $\stau$, 
	eliminating all spin correlations between the taus. Therefore, the probability 
	distribution of $z$ is the integral of \mbox{$F(x_n)\cdot F(x_f)$} over all $x_n$ and 
	$x_f$ with \mbox{$z_n\cdot z_f\le z$}. The probability density can be obtained via 
	the derivative 
	\begin{eqnarray}
		H(z)&=&\frac{\pd}{\pd z}\sqiint\limits_{\genfrac{}{}{0pt}{}{x_n\cdot x_f<z}
		{x_n,x_f<1}}\pd x_n\pd x_fF_n(x_n)\cdot F_f(x_f) \nonumber\\ 
		&=&\left(P_n+P_f\right)\Big[\ln z+2(1-z)\Big]+
		\left(P_n\cdot P_f\right)\Big[4(z-1)-\ln z(4z+1)\Big]-\ln z\;.\qquad 
		\label{EqDiLepFrag}
	\end{eqnarray}
	The di-pion spectra are the probability densities, $N(\mpp)$, of the variable 
	\mbox{$\mpp=\;\sqrt{z}\widehat{m}_{\tau\tau}$}. The probability 
	density $h(\sqrt{z})$ can be obtained from Eq.~(\ref{EqDiLepFrag}) via \mbox{$h(\sqrt{z})=2\sqrt{z}H(z)$}
	and that for $\widehat{m}_{\tau\tau}$ is the linear increasing function $2\widehat{m}_{\tau\tau}$. With that, the pion spectra 
	$N(\mpp)$ can be calculated as
	\begin{eqnarray}
		N(\mpp)&=&4\mpp\int\limits_{\mpp}^{\EP}\frac{1}{\widehat{m}_{\tau\tau}}
		H\left(\frac{\mpp^2}{\widehat{m}_{\tau\tau}^2}\right)\text{d}\widehat{m}_{\tau\tau}
	\end{eqnarray}
	leading to Eq.~\ref{EqPiSpectra}.

\end{appendix}

\bibliography{}

\end{document}